# Analyzing multidimensional movement interaction with generalized cross-wavelet transform

Petri Toiviainen[a,b]* and Martin Hartmann[a,c]

[a]*Finnish Centre for Interdisciplinary Music Research, University of Jyväskylä, Jyväskylä, Finland*

[b]*Department of Music, Art and Culture Studies, University of Jyväskylä, Jyväskylä, Finland*

[c]*Faculty of Information Technology, University of Jyväskylä, Jyväskylä, Finland*

Correspondence details:
    Prof. Petri Toiviainen
    Finnish Centre for Interdisciplinary Music Research
    Department of Music, Art and Culture Studies
    PO Box 35(M)
    40014 University of Jyväskylä
    Finland
    petri.toiviainen@jyu.fi

# Analyzing multidimensional movement interaction with generalized cross-wavelet transform

Humans can synchronize with musical events whilst coordinating their movements with others. Interpersonal entrainment phenomena, such as dance, involve multiple body parts and movement directions. Along with being multidimensional, dance movement interaction is plurifrequential, since it can occur at different frequencies simultaneously. Moreover, it is prone to nonstationarity, due to, for instance, displacements around the dance floor. Various methodological approaches have been adopted to study entrainment, but only spectrogram-based techniques allow for an integral analysis thereof. This article proposes an alternative approach based upon the cross-wavelet transform, a technique for nonstationary and plurifrequential analysis of univariate interaction. The presented approach generalizes the cross-wavelet transform to multidimensional signals. It allows to identify, for different frequencies of movement, interaction estimates of interaction and leader-follower dynamics across body parts and movement directions. Further, the generalized cross-wavelet transform can be used to quantify the frequency-wise contribution of individual body parts and movement directions to overall synchrony. The article provides a thorough mathematical description of the method and includes proofs of its invariance under translation, rotation, and reflection. Finally, its properties and performance are illustrated via examples using simulated data and behavioral data collected through a mirror game task and a free dance movement task.

## 1 Introduction

Humans, as social beings, interact with others routinely and repeatedly. Behavioral coordination between interactants can be observed in a wide variety of activities, such as dancing, music making, sports, and conversation (Shockley & Riley, 2015). Social interaction, which occurs via an alignment of mental attitudes and bodily postures at multiple levels of coordination (Gallotti et al., 2017), can be observed in early childhood and is present throughout lifetime (Richardson et al., 2012). Numerous terms have been used to designate the interdependence of behaviors between interactants (Delaherche et al., 2012); in this article, *coordination*, *synchrony*, and *interaction* are used interchangeably to describe such phenomenon.

    Interpersonal movement coordination, either intentional or not, is a particularly important facet of human interaction, as evidenced initially by communication studies (Cornejo et al., 2017; Lakin, 2013; Latif et al., 2014) and more recently by psychological (Lakin et al., 2003) and psychotherapeutic research (Wiltshire et al., 2020). Due to the intrinsically multimodal nature of human interaction, there are various ways to tackle interpersonal movement coordination from a research perspective. Besides movement-movement interaction, it is possible to study the relationship between movement interaction and other modalities such as audio. Both individual and jointly produced movements and sounds become coupled at multiple temporal scales, for example, in musical performances (Alviar et al., 2020; Jakubowski et al., 2020); these sound-movement synchronies lead, in turn, to integrated perception. Similar interpersonal *entrainment* phenomena are observed in joint dance, where rhythmic coupling emerges through both musical and social bonds (Phillips-Silver et al., 2010).

Although the neural substrates underlying social interaction remain largely unclear (Schilbach et al., 2013), entrainment phenomena might be possible due to auditory-motor and visual-motor neural connections tuned for imitation (Laland et al., 2016). Examples of these connections include auditory-motor interactions in processing of human perceptual information (Gordon et al., 2018; Patel & Iversen, 2014).

Most studies on music-related interpersonal coordination have focused on constrained tasks, such as finger tapping (Repp, 2005; Repp & Su, 2013). This paradigm, which is crucial to sensorimotor synchronization research, has been used to support theoretical accounts of the common coding between perceptual and motor representations (Hommel et al., 2001). Major findings in dyadic tapping research include the identification of *mutual adaptation* coordination phenomena, whereby coupled tappers tend to adjust their behavior in order to continuously adapt to one another rather than exhibiting leader-follower dynamics (Himberg, 2014; Konvalinka et al., 2010). Recent work also illustrated the human propensity for synchronization: in a dyadic tapping task that consisted of maintaining the beat of a metronome, participants unintentionally synchronized with each other regardless of the partner's ability to keep the cued tempo (Schultz & Palmer, 2019). An advantage of tapping as a means to explore the dynamics of human movement interaction is that it allows for rather straightforward analyses; this approach, however, typically has limited ecological validity.

Motion capture systems, whether video-based, optical, electromagnetic, inertial measurement units or otherwise, are widely used in both research and industry. While tapping studies are limited to discrete time series analyses, motion capture techniques allow for the collection of continuous movement data streams. Besides numerous experimental design and analysis possibilities, these techniques allow for collecting precise movements of the human body with relatively high ecological validity.

Dance is a clear example of human movement interaction that can be feasibly investigated under ecological conditions. Dancing with others is considered to serve various social functions, including group bonding, coalition signaling, and courtship (Christensen et al., 2017; Dunbar, 2012). Few studies on dyadic coordination in dance have been conducted, reporting strong relationships between dancers' orientation and observers' ratings of perceived interaction (Carlson et al., 2019; Hartmann et al., 2019). With respect to group coordination, a recent study of folk dancers has shown the effect of haptic coupling (holding hands) and musical coupling (dancing to music) upon horizontal and vertical group synchrony, respectively (Chauvigné et al., 2019). It has also been found that people who were asked to dance in high synchrony exhibited increased self judgements of group bonding (Tarr et al., 2015).

### 1.1 Methods for quantifying movement interaction

Our understanding of human movement coordination and, more generally, of social signals and social behaviors, has been growing rapidly over the last two decades thanks to developments in both optical and computer vision-based human motion capture (Vinciarelli et al., 2009). From a methodological standpoint, it is possible to identify a taxonomy of approaches to interpersonal movement coordination from the following set of dichotomous categories: analysis domain (time-domain/frequency-domain), temporal event type (discrete/continuous), time series stationarity (stationary/non-stationary), and dimensionality (bivariate/multivariate). The applicability of these procedures largely depends on the nature of the data.

*1.1.1 Time-domain methods*

Discrete series of time events, such as tapping times, might represent periodic processes. Several tapping studies, for instance, extract phase angles from tap onset time series in order to compute circular statistics. Directional statistics allow the revelation of complex entrainment patterns while remaining invariant to tempo variation (Caron et al., 2017). One such example is the *synchronization index*, i.e., the mean resultant vector length across unit complex numbers obtained from pairwise differences between instantaneous phases of two time series (Mardia & Jupp, 1990; Tognoli et al., 2007). This measure, which is inversely related to the circular variance of phase differences, has been utilized in various studies to estimate synchronization accuracy either between two tappers or between a tapper and a metronome (Heggli et al., 2019; Konvalinka et al., 2010; Skewes et al., 2014; Witek et al., 2017).

Oftentimes, interpersonal coordination studies investigate leader-follower relationships, that is, whether observations of one individual correspond to temporally lagged observations of the other. This is typically performed via the windowed cross-correlation and peak picking procedure (Boker et al., 2002). In this research area, cross-correlation has been applied to temporal bivariate data comprising both discrete events, such as finger tap times or inter-tap times from a dyad (Repp, 2006). However, the majority of studies have utilized this approach to quantify coordination from continuous events, such as movement in verbal conversations (Galbusera et al., 2018; Ramseyer & Tschacher, 2014; Tschacher et al., 2014, 2018), dance movements (Himberg & Thompson, 2011; Josef et al., 2019), joint musical improvisation features (Luck et al., 2008), and physiological changes (Bar-Kalifa et al., 2019; Tschacher & Meier, 2020). Among its disadvantages are that window and lag range parameters need to be specified, and that, as in correlation, cross-correlation might return misleading coefficients when applied to autocorrelated time series (Dean & Dunsmuir, 2016).

Another issue affecting the study of human movement interaction is that it involves the full body dynamics. Several studies have reduced movement interaction to either a specific marker, such as head or hand movements (Boker et al., 2002; Dotov et al., 2021; Himberg & Thompson, 2011; Josef et al., 2019; Reiss et al., 2019), or computed cross-correlations between few body parts of interest (Ramseyer & Tschacher, 2014). Other approaches (Caramiaux et al., 2009; Godøy et al., 2016; Hartmann et al., 2019) have dealt with multidimensional movement interaction through latent space methods, which are based on fitting a linear subspace to two sets of multivariate data in order to measure their relationship. Canonical Correlation Analysis (CCA, Hotelling, 1936) and Partial Least Squares (PLS, Wold, 1982) methods perform a decomposition of the cross-covariance matrix between two sets of variables in order to return latent variables with maximal correlation or covariance (in CCA or PLS respectively). As in Principal Component Analysis (PCA), these methods require the use of a dimensionality reduction parameter, i.e., the number of linear combinations of the original variables to return. The main advantage of PLS and CCA is their data-driven modelling of interaction, whereby all possible combinations between individuals' body parts –and their movement directions– might contribute to synchronization. Further, these methods ignore whether the movements have identical or opposite phase. For interpersonal movement synchrony research, this becomes a benefit rather than a drawback, since in- and anti-phase relationships seem to constitute the most stable modes of coordination (Miles et al., 2009); further, the prosocial effects of movement synchrony might be driven by its contingency, not by its symmetry (Cirelli et al., 2014; Cross et al., 2016).

A problem common to PLS and CCA is that they carry an implicit assumption of *weak stationarity*. This assumption means that, for a given time series, mean and variance cannot change over time and the autocorrelation of any excerpt of the time series has to depend only on the amount of time by which it has been shifted (Witt et al., 1998). The weak stationarity assumption is often unmet for real movement interaction. For instance, dancers might move more in certain parts of a musical piece than others, which would result in non-stationarity due to changes in variance. Another possible issue is that relevant information about joint movement patterns gets overshadowed by slowly varying, nonstationary trends such as displacements around the dance floor while dancing. While it is possible to overcome nonstationarity to some extent by applying, for instance, a windowed analysis, this approach still assumes that the signals are stationary within each analysis window, a condition that might not necessarily hold true. Furthermore, finding optimal window parameters is not a trivial matter, yet these might have a large impact on the results obtained with PLS and CCA. In particular, small windows can extract temporally local information accurately, but their frequency resolution is poor. By contrast, large windows provide good frequency resolution, but fail to extract temporal fluctuations accurately.

Multidimensional cross-recurrence quantification analysis (Wallot, 2019; Wallot et al., 2016; Wallot & Leonardi, 2018) provides a time-domain method for the analysis of multivariate interaction that does not require windowed analyses. This method can be used to assess the similarity of the joint dynamics of two sets of observations and identify leader–follower relationships (Crone et al., 2021). A limitation of this approach is that both multivariate time series must represent equivalent data dimensions. For example, anteroposterior head movements in the first dimension should be followed by mediolateral head movements, and so forth. Also, cross-recurrence methods do not directly estimate the relative contribution of different variables upon interaction, unlike e.g. latent space methods.

*1.1.2  Frequency-domain methods*

Movement interaction is *plurifrequential*, meaning that it can occur at several frequencies simultaneously (Schmidt et al., 2012; Toiviainen et al., 2010). For individual movement to music (Toiviainen et al., 2010), vertical movements seem to commonly occur at a frequency of about 2 Hz (i.e., at every beat), whereas sideways swaying tends to occur at 0.5 Hz (i.e., every four beats). Since time-domain methods do not decompose the time series into different frequency bands, they may not be sensitive enough to detect nested periodicities. Furthermore, in certain research areas, finding at which frequency –or frequencies– the interaction occurs can be of interest per se. A common frequency-domain method to quantify interaction is Cross-spectral analysis, which is based on the pointwise multiplication of two complex-valued Fourier transforms (Schmidt et al., 2012). However, since this method assumes stationarity in the data, pointwise multiplication of two short-time Fourier transforms (STFTs) needs to be calculated for nonstationary data.

An alternative to deal with nonstationarity in the frequency domain is to perform a scale-independent time-frequency localization. *Cross-wavelet transform* (XWT) is similar to the cross-spectrogram technique, but uses a discrete wavelet transform (DWT) instead of a STFT to obtain the time-frequency representations of the signals (Hudgins et al., 1993; Issartel et al., 2014). The DWT decomposes a time series into components that are localized in time and frequency by performing a number of convolutions between a time series and dilations/contractions of a function called mother wavelet. Since each wavelet (i.e., each dilation and contraction of the mother wavelet) represents a different frequency interval in the frequency domain, wavelets are comparable to bandpass filters (Hudgins et al., 1993). A complex wavelet function returns not only amplitude, but also phase information.

The time-frequency representation obtained with DWT, and thus XWT, differs from that of the STFT in the following aspect: while the STFT has a fixed temporal and frequency resolution, the wavelet transform yields high temporal resolution for high frequency channels and high frequency resolution for low-frequency channels. XWT analysis is similar to cross-spectrogram analysis using STFTs, but with the added advantages of variable time-frequency localization and accurate phase lag estimates for each frequency channel.

Numerous recent studies have studied movement interaction through cross-wavelet analysis (Clayton et al., 2019; Dotov et al., 2021; Eerola et al., 2018; Fujiwara et al., 2020; Fujiwara & Daibo, 2016; Jakubowski et al., 2020; Walton et al., 2015; Wiltshire et al., 2019). Although substantial descriptors can be computed from the wavelet cross-spectrum, including magnitude-squared coherence and phase difference information, this method is limited to bivariate analysis only. This approach is appropriate for the study of various kinds of low-dimensional interpersonal movement coordination, such as body sway in sitting musicians, but it fails to accurately describe full-body forms of interaction such as dance.

We present a generalization of the cross-wavelet transform that allows the quantification of multidimensional, multifrequency, and nonstationary movement interaction. The method is based on performing a XWT on each pair of movement data dimensions followed by evaluating, for each time-frequency point, the global amplitude and phase relation of the two multidimensional time series by estimating the distribution of the pairwise XWTs in the complex plane. While a few implementations of the problem have been proposed in other research domains (Chavez & Cazelles, 2019; Soon et al., 2014), the method proposed in the present article is particularly suitable for movement analysis, because it is translation-, rotation-, and reflection-invariant and thus not dependent on the location, orientation and handedness of the used coordinate system.

In the remainder of this article, we will first give a detailed explanation of the generalized cross-wavelet transform for analysis of multidimensional movement interaction, including proofs of its invariance under different transformations—translation, rotation, and reflection. Next, we will illustrate some properties of the proposed method. First, using simulated data (sets of sinewaves), we (1) derive mathematical approximations for the phase and amplitude of the generalized cross-wavelet transform. Second, using motion capture position data collected from a standardized mimicry task and a free dance movement study we show that the transform (2) allows the identification of leader-follower relationships at different frequencies of movement; (3) offers frequency-wise estimates of interaction for different body parts and movement directions; and (4) can be used to quantify, separately for each frequency, the contribution of individual body parts and movement directions to the overall synchrony.

## 2   Method

Conceptually, the generalized cross-wavelet transform (henceforth GXWT) is based on the calculation of bivariate cross-wavelet transforms between all possible pairs of individual signal components in **X** and **Y**, thus yielding a total of $N \times M$ cross-wavelet transforms. Subsequently, for each point in the time-frequency plane, the distribution of the thus obtained $N \times M$ cross-transform values is modelled with a bivariate normal distribution in the complex plane. The size and shape of the estimated distribution, more specifically, the variance and eccentricity thereof, provide an aggregate measure of the degree of time- and frequency-localized synchrony between the multivariate time series. In particular, a distribution with a large major axis and high eccentricity indicates strong synchrony, while a more circular distribution indicates weaker synchrony. On the other hand, the angle between the distribution's major axis and the real axis of the complex plane provides an aggregate measure of the mutual phase difference between the multivariate time series. Matlab codes implementing the GXWT algorithm are provided as supplementary material (https://doi.org/10.17011/jyx/dataset/75063).

In what follows, a mathematical description of the method is provided.

(1) Let $\boldsymbol{X}$ and $\boldsymbol{Y}$ denote multidimensional time series with dimensions of $N$ and $M$, respectively, and with $T$ time points:

$$\boldsymbol{X} \in \mathbb{R}^{T \times N}, \boldsymbol{Y} \in \mathbb{R}^{T \times M} \qquad (1)$$

(2) By applying discrete wavelet transform $\mathfrak{W}(\cdot)$ to each of the time series components in turn we obtain complex-valued wavelet transform tensors

$$\boldsymbol{\mathcal{U}} := \mathfrak{W}(\boldsymbol{X}) \in \mathbb{C}^{F \times T \times N}, \boldsymbol{\mathcal{V}} := \mathfrak{W}(\boldsymbol{Y}) \in \mathbb{C}^{F \times T \times M} \qquad (2)$$

where $F$ is the number of frequency channels.

(3) At each time-frequency point $(f,t)$, the tensor fibers $\boldsymbol{u}_{ft}$ and $\boldsymbol{v}_{ft}$,

$$\boldsymbol{u}_{ft} := \boldsymbol{\mathcal{U}}(f,t,:) \in \mathbb{C}^{N \times 1}, \boldsymbol{v}_{ft} := \boldsymbol{\mathcal{V}}(f,t,:) \in \mathbb{C}^{M \times 1} \qquad (3)$$

represent the multivariate time-frequency-localized amplitude and phase of $\boldsymbol{X}$ and $\boldsymbol{Y}$, respectively.

(4) Time-frequency localized cross-spectrum matrix at $(f,t)$ is obtained via the outer product of $\boldsymbol{u}_{ft}$ and $\boldsymbol{v}_{ft}^*$:

$$\boldsymbol{M}_{ft} = (m_{jk})_{ft} := \boldsymbol{u}_{ft} \boldsymbol{v}_{ft}^{\star T} \in \mathbb{C}^{N \times M} \qquad (4)$$

The matrix $\boldsymbol{M}_{ft}$ thus contains time-frequency localized cross-wavelet transform values for all possible pairs between the individual time series in $\boldsymbol{X}$ and those in $\boldsymbol{Y}$.

(5) The degree of circular asymmetry of the distribution of values $(m_{jk})_{ft}$ in the complex plane is a measure of time-frequency localized synchrony between $\boldsymbol{X}$ and $\boldsymbol{Y}$. To this end, we model the values $(m_{jk})_{ft}$ as a zero-mean complex normal distribution:

$$(m_{jk})_{ft} \sim \mathcal{CN}(0, \Gamma_{ft}, C_{ft}) \qquad (5)$$

where $\Gamma_{ft}$ denotes variance and $C_{ft}$ pseudo-variance. The size, shape and orientation of the estimated normal distribution are taken as measures of synchrony. In particular, the elongation of the normal distribution indicates the strength of synchrony, while the orientation indicates the phase shift. It can be shown (Ollila, 2008) that these can be estimated by sample pseudo-variance

$$\tau_{ft} := \frac{1}{N^2} \sum_{jk} (m_{jk})_{ft}^2 \in \mathbb{C} \tag{6}$$

In particular, the modulus of the sample pseudo-variance equals the product of the sample variance $S_{ft}^2$ and squared eccentricity of the estimated distribution

$$|\tau_{ft}| = S_{ft}^2 \varepsilon^2 \tag{7}$$

thus providing a measure combining the size and elongation of the estimated distribution. The orientation angle between the distribution's major axis and the real axis, $\alpha$, can be obtained by:

$$\alpha = \frac{arg(\tau_{ft})}{2} \tag{8}$$

Figure 1 shows an example of distribution of instantaneous cross-wavelet transform values and the estimated distribution.

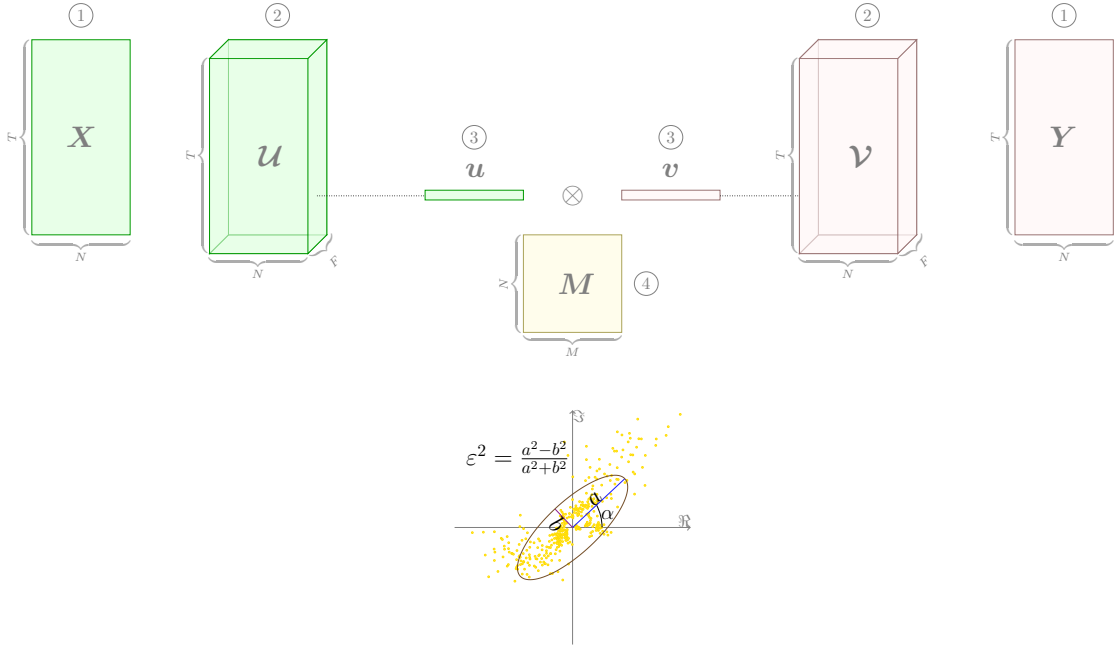

Figure 1. Top: Schematic diagram of the method, covering steps 1-4 of its mathematical description. Bottom: Example distribution in the complex plane of the elements $(m_{ij})_{ft}$ of a time-frequency localized cross-spectrum matrix $M_{ft}$. The ellipse shows the 95% confidence region of the estimated complex normal distribution $\mathcal{CN}(0, \Gamma_{ft}, C_{ft})$.

(6) The matrix

$$C := (c_{ft}) = (\sqrt{\tau_{ft}}) \in \mathbb{C}^{F \times T} \tag{9}$$

where $\sqrt{\cdot}$ refers to the principal square root, constitutes the GXWT. With this definition, the real part of GXWT is identically non-negative and consequently $-\pi/2 \leq arg(c_{ft}) \leq \pi/2$.

(7) The contribution of each channel pair in $X$ and $Y$ to time-frequency localized global interaction can be estimated by the location of the respective component in relation to the estimated elliptical distribution. In particular, if an element resides close to the distributions' major axis and distant from the origin, its contribution is considered high. A numerical estimate thereof can be obtained by projecting each element of $M_{ft}$ onto the major axis of the estimated distribution, and taking the absolute value of the projection:

$$P_{ft} = (p_{jk})_{ft} := \left\| \Re(M_{ft} e^{-i\arg(c_{ft})}) \right\| \tag{10}$$

Subsequently, the time-frequency localized contribution of each channel $X$ and $Y$, can be estimated by taking the row- and column-wise means of $P_{ft}$

$$(p_X)_{ft} := \frac{1}{M} P_{ft} j_M \in \mathbb{R}_+^N \tag{11}$$

$$(p_Y)_{ft} := \frac{1}{N} P_{ft}^T j_N \in \mathbb{R}_+^M \tag{12}$$

where $j_M \in \mathbb{R}^{M \times 1}$ is a column vector of ones. Performing this for each time-frequency point we obtain the real-valued projection tensors

$$P_X := (p_X)_{ft} \in \mathbb{R}_+^{F \times T \times M}, \quad P_Y := (p_Y)_{ft} \in \mathbb{R}_+^{F \times T \times M} \tag{13}$$

which provide estimates of the contribution of each signal channel to global interaction at each time-frequency point.

(8) If $X$ and $Y$ have the same dimensionality $N > 1$ and the respective dimensions in them represent equivalent data (e.g., the same body part and spatial dimension in motion capture data of two different persons), it may be in some cases useful to consider only pairwise synchronization between respective dimensions. In this case the time-frequency localized cross-transform can be obtained via the sample pseudo-variance $\tau_{ft}^\circ$ of the elements of the Hadamard product $m_{ft}^\circ = (m_j^\circ)_{ft} := u_{ft} \circ v_{ft}^\star$:

$$\tau_{ft}^\circ := \frac{1}{N} \sum_j (m_j^\circ)_{ft}^2 = \frac{1}{N} (u_{ft} \circ u_{ft})^T (v_{ft}^\star \circ v_{ft}^\star) \tag{14}$$

(9) Relationship with bivariate cross-wavelet transform: when $X$ and $Y$ are univariate, we have $u_{ft} = u_{ft} \in \mathbb{C}^1$, $v_{ft} = v_{ft} \in \mathbb{C}^1$, and by denoting the bivariate cross-wavelet transform at time-frequency point $(f,t)$ by $\hat{c}_{ft}$, we get

$$c_{ft} = \sqrt{u_{ft}^2 v_{ft}^{\star 2}} = \pm u_{ft} v_{ft}^\star = \pm \hat{c}_{ft} \tag{15}$$

and thus

$$\begin{cases} |c_{ft}| = |\hat{c}_{ft}| \\ \arg(c_{ft}) = \arg(\hat{c}_{ft}) \mod \pi \end{cases} \tag{16}$$

Consequently, the generalized cross-wavelet transform reduces to bivariate cross-wavelet transform with the exception that the former does not make a distinction between in-phase and anti-phase movement. This is compatible with multivariate time-domain methods such as Canonical Correlation Analysis and Partial Least Squares Correlation in the sense that these methods, being covariance-based, ignore the sign of correlation in their latent space projections.

## 2.1 Transformation invariances

(10) With data representing three-dimensional movement in a Cartesian coordinate system, such as obtained with motion capture, GXWT is invariant under translation, rotation, and reflection, thus being independent on the particular location, orientation, and handedness of the used coordinate system.

*Translational invariance* follows from the linearity of the wavelet transform. In particular, for any signal $x$ and any constant signal $C$ it holds
$$\mathfrak{W}(x + C) = \mathfrak{W}(x) + \mathfrak{W}(C) = \mathfrak{W}(x) \tag{17}$$

To prove *rotational invariance*, let **X** and **Y** be three-dimensional time series
$$\mathbf{X}, \mathbf{Y} \in \mathbb{R}^{3 \times T} \tag{18}$$

and let
$$\mathbf{Z} = \mathbf{R}\mathbf{X} \tag{19}$$

where $\mathbf{R} \in \mathbb{R}^{3 \times 3}$ is a rotation matrix. Then, due to linearity of the wavelet transform,
$$\mathbf{\mathcal{W}} := \mathfrak{W}(\mathbf{Z}) = \mathfrak{W}(\mathbf{R}\mathbf{X}) = \mathbf{R}\mathfrak{W}(\mathbf{X}) \tag{20}$$

and thus
$$\mathbf{w}_{ft} := \mathbf{\mathcal{W}}(:,f,t) = \mathbf{R}\mathbf{u}_{ft} \tag{21}$$

Given that for any matrix $\mathbf{A} = (a_{jk})$ we have the equivalence $\sum_{jk} a_{jk}^2 \equiv tr(\mathbf{A}^T\mathbf{A})$, the sample pseudo-variance $\tau_{ft}$ of Eq. 6 can be written in matrix notation as follows:
$$\tau_{ft} := \tfrac{1}{N^2}\sum_{jk}(m_{jk})_{ft}^2 = \tfrac{1}{N^2}tr(\mathbf{M}_{ft}^T\mathbf{M}_{ft}) = \tfrac{1}{N^2}tr(\mathbf{v}_{ft}^\star \mathbf{u}_{ft}^T \mathbf{u}_{ft} \mathbf{v}_{ft}^{\star T}) \tag{22}$$

Therefore, the pseudo-variance between **Z** and **Y**, denoted by $\widetilde{\tau_{ft}}$, is
$$\widetilde{\tau_{ft}} = \tfrac{1}{N^2}tr(\mathbf{v}_{ft}^\star \mathbf{w}_{ft}^T \mathbf{w}_{ft} \mathbf{v}_{ft}^{\star T}) = \tfrac{1}{N^2}tr(\mathbf{v}_{ft}^\star \mathbf{u}_{ft}^T \mathbf{R}^T \mathbf{R} \mathbf{u}_{ft} \mathbf{v}_{ft}^{\star T}) = \tau_{ft} \tag{23}$$

due to the general property $\mathbf{R}^T\mathbf{R} \equiv \mathbf{I}$ of rotation matrices.

With $N = 3K$ and the same rotation applied to all K groups of 3 time series, the rotation matrix $\mathbf{R}$ is a block diagonal matrix of $3 \times 3$ rotation matrices, for which $\mathbf{R}^T\mathbf{R} \equiv \mathbf{I}$ again applies generally and thus rotation invariance holds. The rotational invariance does not hold for the pair-wise coupling scheme presented in item 9.

Finally, the GXWT is *reflection invariant*. This can be seen by first observing that any reflection can be reduced to a combination of translation, rotation, and reflection across one or several coordinate axes. The latter is equivalent to multiplying the respective signal components by -1, and invariance under this transform follows from the linearity of the wavelet transform and the even symmetric property of the pseudovariance as a function of $(m_{jk})$ (Eq. 6): if any of the components $m_{jk}$ is multiplied by –1, the sample pseudovariance does not change.

# 3 Examples

## 3.1 Example 1: Simulated data

Let $X$ and $Y$ be multivariate stationary signals each consisting of $N$ sinewaves of amplitudes $A_x$ and $A_y$ respectively and frequency $f_0$ with normally distributed phase differences

$$X(t) = A_x cos(f_0 t + \alpha_x + \boldsymbol{\beta_x}), \boldsymbol{\beta_x} = [\beta_{x1}, \ldots, \beta_{xN}], \beta_{xk} \sim \mathcal{N}(0, \sigma_x^2)$$

$$Y(t) = A_y cos(f_0 t + \alpha_y + \boldsymbol{\beta_y}), \boldsymbol{\beta_y} = [\beta_{y1}, \ldots, \beta_{yN}], \beta_{yk} \sim \mathcal{N}(0, \sigma_y^2)$$

See Figure 2 for an example of such signals.

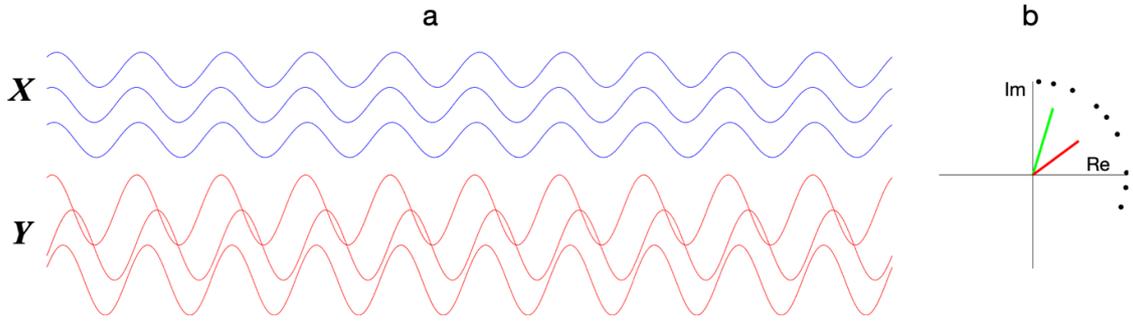

Figure 2. (a) Signals $X$ and $Y$ generated with parameters $N = 3$, $f_0 = 1\ Hz$, $\alpha_x = 0$, $\alpha_y = \pi/2$, $A_x = 2$, $A_y = 1$, $\sigma_x^2 = 0.25$, $\sigma_y^2 = 0.5$; (b) time-frequency localized cross-transform values $(m_{jk})_{f_0 t}$ in complex plane (black dots), together with pseudo-variance $\tau_{f_0 t}$ (green line), and square root of pseudo-variance $\sqrt{\tau_{f_0 t}}$ (red line).

Using analytic wavelets, the components of their wavelet transforms corresponding to frequency $f_0$ are the $N$-dimensional analytic signals

$$\boldsymbol{u}_{f_0 t} = A_x e^{i(f_0 t + \alpha_x + \boldsymbol{\beta_x})}$$

$$\boldsymbol{v}_{f_0 t} = A_y e^{i(f_0 t + \alpha_y + \boldsymbol{\beta_y})}$$

where $i$ is the imaginary unit. The components of the time-frequency localized cross-spectrum matrix $\boldsymbol{M}_{f_0 t}$ are

$$(m_{jk})_{f_0 t} = A_x A_y e^{i(\alpha_x - \alpha_y + \beta_{xj} - \beta_{yk})} = A_x A_y e^{i(\alpha + \delta_{jk})},$$

where

$$\alpha = \alpha_x - \alpha_y$$

is the mean pairwise phase difference between components of $X$ and components of $Y$, and

$$\delta_{jk} = \beta_{xj} - \beta_{yk} \sim \mathcal{N}(0, \sigma_x^2 + \sigma_y^2).$$

It must be noted that due to stationarity $(m_{jk})_{f_0 t}$ are constant over time. The pseudo-variance of $\boldsymbol{M}_{f_0 t}$ is

$$\tau_{f_0 t} = \frac{A_x^2 A_y^2}{N^2} \sum_{jk} e^{2i(\alpha + \delta_{jk})} = e^{2i\alpha} \frac{A_x^2 A_y^2}{N^2} \sum_{jk} e^{2i\delta_{jk}}$$

Since for the expectation value of the sum term it holds

$$\arg(\langle \sum_{jk} e^{2i\delta_{jk}} \rangle) = 0,$$

we get

$$arg\left(\sqrt{\langle \tau_{f_0 t} \rangle}\right) = \begin{cases} \alpha, & 0 < |\alpha| < \pi/2 \\ \alpha + \pi, & \pi/2 < |\alpha| < \pi \end{cases}$$

and thus the argument of the square root of $\langle \tau_{f_0 t} \rangle$ is equal to the mean pairwise difference, modulo $\pi$, between the components of $\boldsymbol{X}$ and $\boldsymbol{Y}$. Thus, similarly to correlation- and covariance-based methods such as CCA and PLS, the generalized cross-wavelet transform does not make any difference between in-phase and anti-phase signals.

For small phase dispersion, $\sigma_x^2 + \sigma_y^2 \ll 1$, using the approximation

$$e^x \approx 1 + x + \frac{x^2}{2}$$

we get for the expectation value of the pseudo-variance

$$\begin{aligned}
\langle \tau_{f_0 t} \rangle &\approx e^{2i\alpha} \frac{A_x^2 A_y^2}{N^2} \langle \sum_{jk} (1 + 2i\delta_{jk} - 2\delta_{jk}^2) \rangle \\
&= e^{2i\alpha} \frac{A_x^2 A_y^2}{N^2} (N^2 + 2i\langle \sum_{jk} \delta_{jk} \rangle - 2\langle \sum_{jk} \delta_{jk}^2 \rangle) \\
&= e^{2i\alpha} \frac{A_x^2 A_y^2}{N^2} [N^2 + 0 - 2N^2(\sigma_x^2 + \sigma_y^2)] \\
&= e^{2i\alpha} A_x^2 A_y^2 [1 - 2(\sigma_x^2 + \sigma_y^2)].
\end{aligned}$$

Therefore, for small phase dispersion

$$\left|\sqrt{\langle \tau_{f_0 t} \rangle}\right| \approx A_x A_y \sqrt{1 - 2(\sigma_x^2 + \sigma_y^2)} \approx A_x A_y (1 - \sigma_x^2 - \sigma_y^2),$$

and thus for zero phase dispersion the square root of pseudo-variance equals the product of the amplitudes, and for small non-zero phase dispersion this value reduces by a factor equal to the total phase dispersion $\sigma_x^2 + \sigma_y^2$.

### *3.2 Example 2: Motion capture data from a mirror game*

The data used in this example was collected with an optical motion capture device during a "mirror game" task (see e.g. Feniger-Schaal et al., 2020), in which two participants were instructed to mirror the hand movement of their partner. The data thus consisted of three dimensions per participant. The experiment consisted of three conditions. In conditions 1 and 2, participants 1 and 2 were instructed to lead, respectively, while in condition 3 none of the participants was instructed to lead.

Figure 3 shows motion capture from a selected dyad performing the task during condition 3. Additionally, it displays the wavelet transforms and the derived cross transform at the selected frequency of f = 0.9 Hz during the three conditions. As is evident from panels c and d of the figure, during 0-4 seconds in the segment there is virtually no movement at this frequency, during 4-8 seconds participant 1 has the lead (as indicated by the positive imaginary part of the cross transform at this frequency), while during 8-12 seconds the roles are reversed (negative imaginary part).

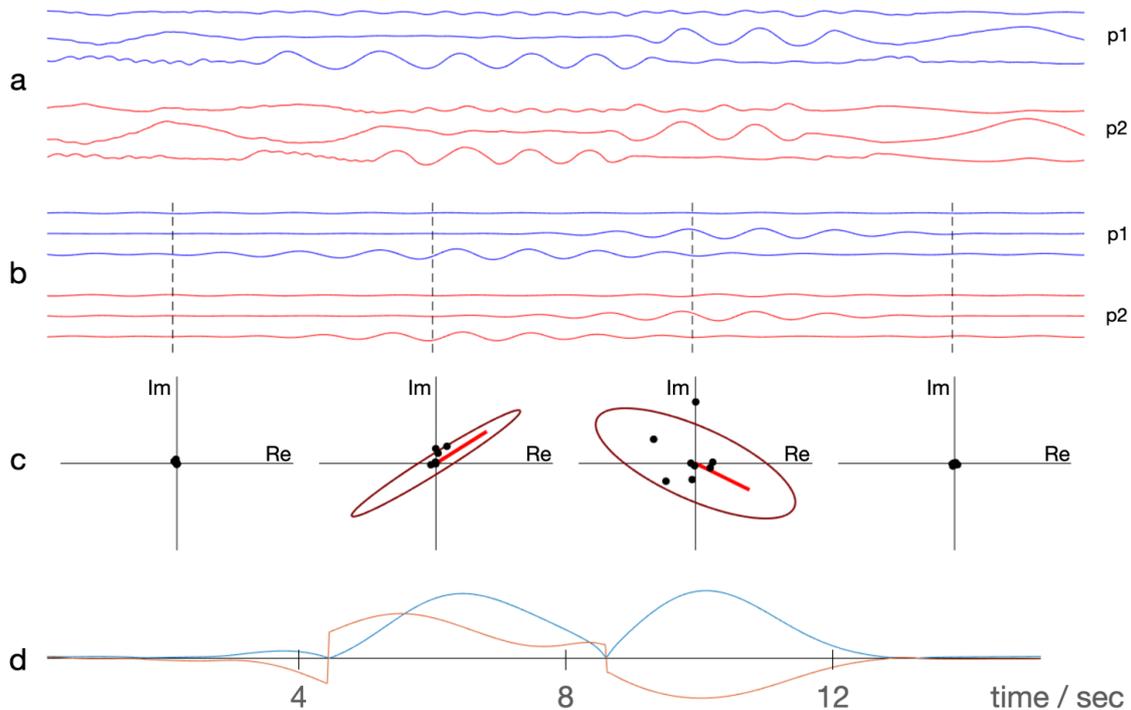

Figure 3. (a) Sixteen seconds of movement data of one hand for two participants (p1 and p2) performing a mirror game task. The three dimensions of each data correspond, from top to bottom, to mediolateral, anteroposterior, and vertical movement. (b) Real part of wavelet transform for each data dimension at the frequency of 0.9 Hz. (c) Cross-transform matrix values $(m_{jk})_{ft}$ (black dots) for time points 2, 6, 10, and 14 seconds (dashed lines), isodensity ellipses at 95% confidence regions of the respective estimated distributions, and square root of pseudo-variance $\sqrt{\tau_{ft}}$ (red lines). (d) Real (blue) and imaginary (red) part of the generalized cross-transform at the selected frequency.

Figure 4 displays the real and imaginary parts of the generalized cross-spectra for another dyad for three conditions: participant 1 leading, participant 2 leading, and none of participants leading. As is evident from the imaginary parts, for conditions 1 and 2 the interaction was strongest at the frequency range of 0.25 ... 0.3 Hz and the phase difference was in accordance with the instruction. In the third condition, the interaction happened at a lower frequency of ca. 0.125 Hz, and the imaginary was close to zero, indicating near zero-phase or anti-phase locking.

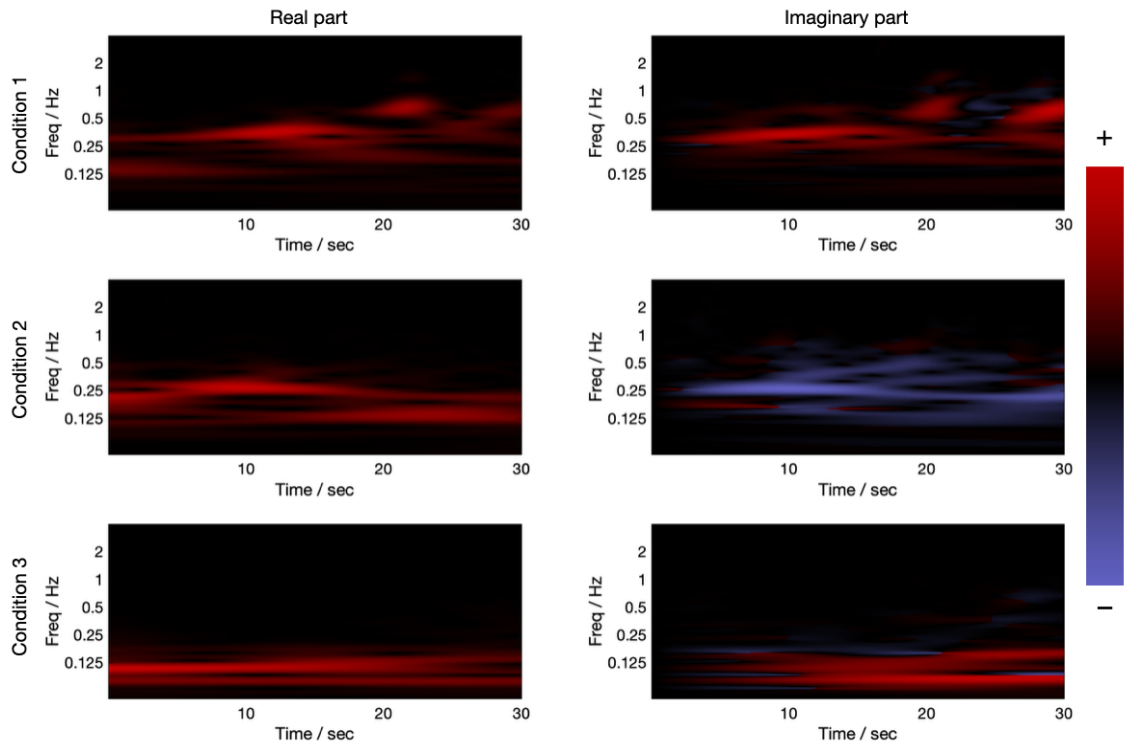

Figure 4. Real and imaginary parts of generalized cross-wavelet transforms from motion capture data of two participants performing a mirror game with hand movement. Condition 1: participant 1 leading; condition 2: participant 2 leading; condition 3: none of the participants leading.

## 3.3 Example 3: Perceived interaction during spontaneous dance to music

This example is based on data published in Carlson, Burger & Toiviainen (2019). They collected motion capture data of 24 dyads spontaneously dancing to pieces of pop music. Participants' movements were recorded using a twelve-camera optical motion capture system (Qualisys Oqus 5+), tracking at a frame rate of 120 Hz, the three-dimensional positions of 21 reflective markers attached to each participant. The locations of the markers are shown in Figure 5.a. Following this, the data were transformed into a set of 20 secondary markers for each dyad member, as shown in Figure 5.b. After the transformation, the motion capture data for each dancer comprised 60 channels (20 markers with three dimensions each).

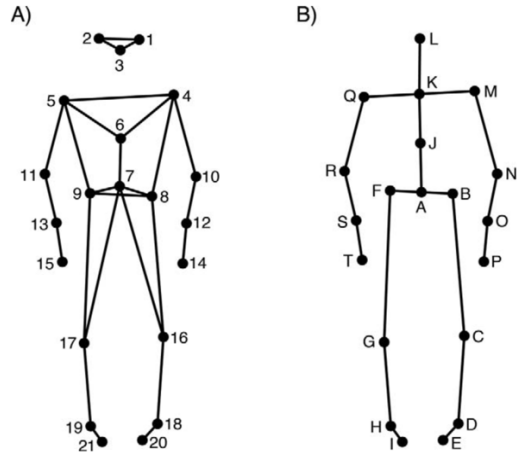

Figure 5. (a) Primary and (b) secondary markers used in the example study.

Subsequently, stick figure animations generated from the motion capture data were presented to participants during a perceptual study, in which the task was to rate perceived interaction in the dance movements. For further details of the data collection, please refer to Carlson, Burger & Toiviainen (2019).

For the purpose of this example, generalized cross-wavelet transform analysis was subsequently performed for five channel groups: all channels (60 per dancer), vertical channels (20), horizontal channels (40), hands (6), and feet (6). Subsequently, for each channel group, the *interaction spectrum* was obtained by calculating the frequency-wise temporal mean of the modulus of the transform according to

$$|c|_f = \frac{1}{T}\sum_t |c_{ft}|.$$

Following this, the interaction spectra were correlated frequency-wise with the interaction ratings, using Spearman correlation. The results are displayed in Figure 6. As can be seen, hand movements correlate most highly with the perceptual ratings, in particular for frequencies above 1 Hz, peaking at slightly above 1 Hz with correlation $r(22) = .75$. Moreover, vertical movement interaction correlates highly with perceived interaction at frequencies above 2 Hz, the same holding for horizontal interaction at the frequencies above 1 Hz. In contrast, foot movement interaction does not correlate to a high degree with perceived interaction at any frequency.

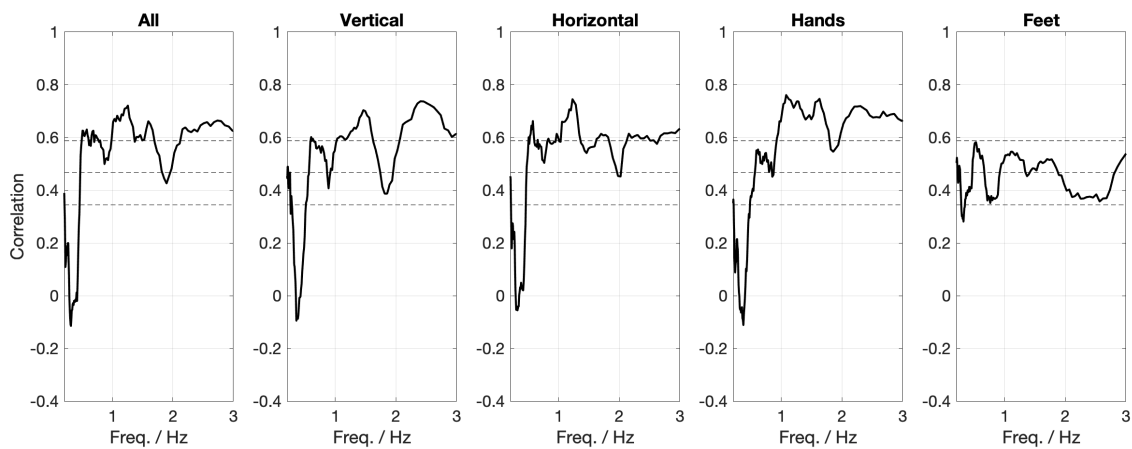

Figure 6. Frequency-wise Spearman correlations between ratings of perceived interaction of 24 dancing dyads and temporal means of cross-wavelet transforms for selected motion capture channel groups. Dashed lines indicate, from bottom to top, the p<.05, p<.01, and p<.001 significance levels (uncorrected for multiple comparisons).

### 3.4 Example 4: Contribution of individual channels to multivariate synchrony

This example is based on the same data as Example 3. The head marker data from one dyad was analyzed using the GXWT. The dancers were facing each other, and the analyzed movement data consisted of three channels for each participant, corresponding to mediolateral, anteroposterior and vertical movement. Subsequently, the contributions of each channel to global head synchrony were calculated according to Eqs. 10-12. The results are displayed in Figure 7.

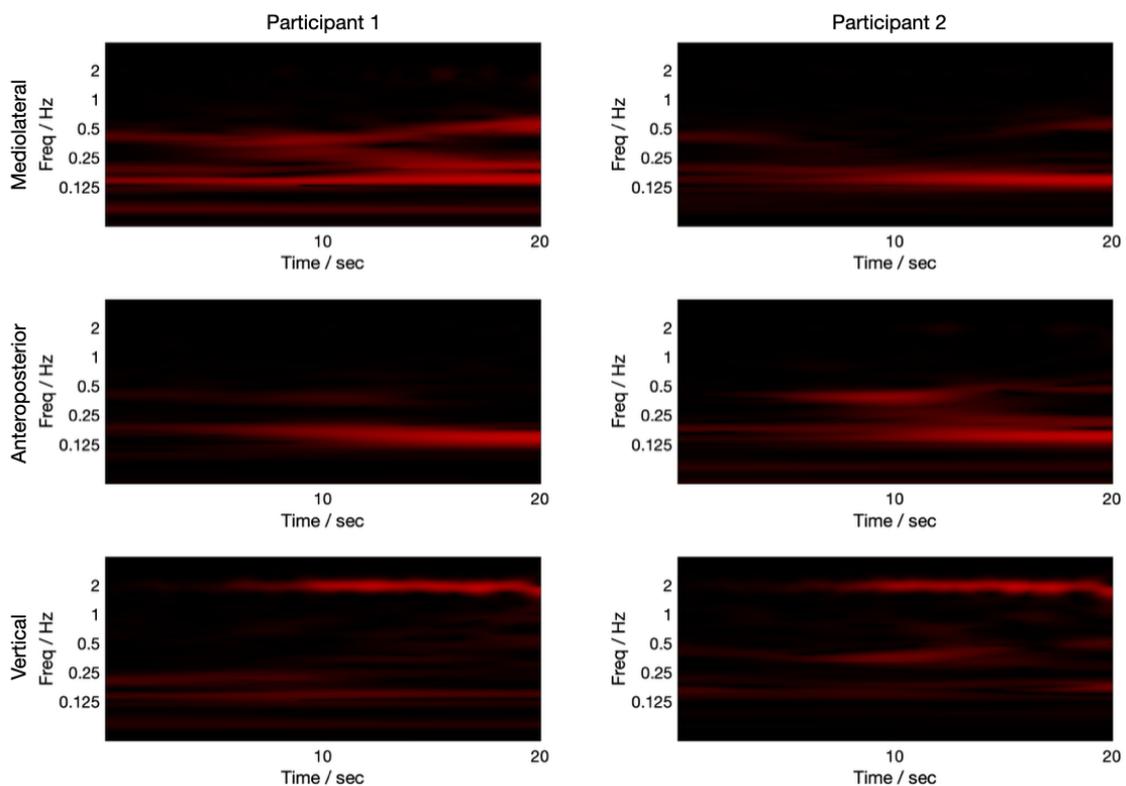

Figure 7. Contributions of mediolateral, anteroposterior and vertical head movement to inter-dancer interaction as a function of time and frequency.

As can be seen, mediolateral and anteroposterior interaction resides predominantly within the frequency range of 0.125 ... 0.5 Hz, while for the vertical movement the predominant interaction frequency is around 2 Hz. This is in agreement with results from previous work that has shown that spontaneous dance movement entails a hierarchical structure, with vertical movement being predominantly synchronized with the basic pulse of music typically having a frequency of ca. 2 Hz, while horizontal movements are synchronized with slower pulse levels (Burger et al., 2018; Toiviainen et al., 2009).
   This approach also allows the identification of differences between dancers in the way that individual movements contribute to the global interaction. Here, a main difference between dancers can be observed at about 0.5 Hz. For participant 1, this interaction is mainly mediolateral: at that frequency, and for most time points, the mean interaction between mediolateral head movement of participant 1 and all head movements of participant 2 is relatively high. In contrast, for Participant 2, the interaction at around 0.5 Hz tends to be anteroposterior, although more distributed among all directions.

# 4 Discussion

In this article, we presented a generalization of wavelet cross-spectrum to the multivariate case. The proposed method allows for quantification of plurifrequential modes of interaction from continuous, non-stationary, multivariate data. Among its advantages, the approach does not distinguish in-phase from anti-phase locking between time series and it is invariant under various data transformations. Furthermore, pairs of time series need not have the same number of dimensions. The latter point is noteworthy, since dimensions exhibiting a significant amount of missing time points – for instance, in pose estimation data– can be discarded without compromising the analysis in some cases.

    The examples above have illustrated the potential of the GXWT as an approach to quantify interpersonal coordination. Furthermore, the proposed method is equally applicable to both social cognition with and without mutual alignment (Gallotti et al., 2017). For instance, the first two conditions in Example 2, in which one of the participants was leading, can be understood as cases of social cognition without mutual alignment since the follower is instructed to align to the leader, whereas the leader is instructed not to adjust to the follower. In contrast, condition 3, in which none of the participants was leading, would correspond to social cognition with mutual alignment: either both participants exchange information reciprocally and adapt to each other, or, due to individual differences, one of the partners adapts more than the other one. The dancing data used for Examples 3 and 4 could be understood as instances of social cognition which may be without mutual alignment, due to reliance on information from a predictable musical pulse. However, due to the unconstrained nature of the task, mutual alignment might be observed depending on the degree of social entrainment between dancers. On a more general note, the GXWT can help describe the types of social interaction that are in play, e.g., by revealing forms of alignment between subjects that do not seem to be associated to periodicities in the music or other external factors.

    A caveat—which also applies to the cross-wavelet transform—must be given regarding leader-follower relationships, such as those described in Example 2. While the averaging of magnitudes across time and frequency is a valid approach to yield a single measure of interaction (see, e.g., Dotov et al., 2021), phase difference analyses should not be averaged across frequency, since different frequencies can yield incompatible information. For instance, the phase difference between a wavelet transform and a temporally delayed copy of it will be larger for higher frequency channels, which have a larger temporal resolution. Lower cross-transform frequencies, in contrast, are sensitive to leader-follower relationships regarding slower movements. Exceptionally, averaging across neighboring frequencies can reveal estimates of phase difference that account for small temporal instability in interaction patterns.

    Related to this, tempo drifts in the data should be taken into consideration when computing an interaction spectrum. This measure carries the implicit assumption of stability of interaction frequencies over time, which is met, for instance, in Example 3

because the dancers are moving to music with a constant tempo. In contrast, whenever the driving stimulus has an irregular beat, the stationarity assumption does not hold and additional analytical steps might be required, such as time alignment or tracking of multiple time-varying frequency components in the wavelet cross-scalogram (Capizzi et al., 2020; Vuoskoski et al., 2014).

On a more general note, methods based on wavelet transform suffer from boundary effects (Torrence & Compo, 1998); these can be rather serious for short time series (i.e., data with a relatively small number of samples), particularly at lower frequency channels. This should be taken into consideration, for instance, when computing the interaction spectrum for lower frequencies of the GXWT. It is possible to either discard the lower frequency part of the interaction spectrum or to compute it only from time-frequency points located within the cone of influence of the GXWT in order to account for these edge effects.

A number of variables can affect the computational speed of the GXWT, including the number of target frequency components and the length and dimensionality of the data. It is possible, however, to obtain the GXWT for few specific target frequencies (such as 2 Hz); this can greatly increase the computational time and cost of the method. Other possibilities for speedup include downsampling the data or applying a dimensionality reduction technique –such as PCA– separately to each data set before calculating the GXWT.

For certain types of data, it is possible to compute pairwise synchronization between respective dimensions via the GXWT. Pairwise synchronization can be used whenever both sets of data have the same number of dimensions, and might be useful, for instance, in analyzing the amount of synchrony in coordinated dances. Besides this possibility, there are various ways in which the GXWT can be modified to better suit certain types of data. Besides the choice of the mother wavelet and a number of wavelet transform parameters, it is possible to apply weighting to the time series dimensions, which will have an effect upon the GXWT.

While we have paid special attention to the applicability of the GXWT for dyadic interaction analysis, this method can also be used e.g. for quantification of dance style by calculating the GXWT between sets of body parts and/or directions of single dancers, or to investigate relationships between individual movements and music danced to at different frequencies. For instance, it is possible to estimate synchrony and phase differences between left and right limbs (that is, bilateral (anti)symmetry), or between vertical and horizontal movements of individuals. Further, both inter- and intra- individual multivariate synchrony analyses are also appropriate in other application domains, including psychotherapy, where interpersonal movement coordination has seen a broad interest due to its potential to improve psychotherapeutic research and practice (Wiltshire et al., 2020). Due to its flexibility regarding differences in dimensionality of the input data, the approach can be used to quantify synchrony between different types of agents, such as humans and animals or humans and robots: it is possible to compute the GXWT even when a different number of, for instance, motion capture markers would be needed for each of the two sets of data.

Moreover, in GXWT, the dimensions of a multivariate time series do not need to be related to the same person. The method can be applied to analyze group-level or collective movement patterns at both 'intra' and 'inter' levels. For example, it can be used to quantify interaction between body parts and/or movement directions for a group of dancers, to estimate synchrony between musicians and audience during concerts or between two species of free-ranging animals, and so on. Future directions for movement interaction include an extension of the GXWT to multiple agents. This might be possible, for instance, via calculation of outer product of multiple vectors and of circular variance to get real and imaginary parts, respectively (Cong et al., 2015).

Although this article has focused on the GXWT as a promising approach for research on movement interaction, this method has wide range applicability. The presented technique can be highly useful for the study of relationships between sets of simultaneous time series, such as those involving continuous perceptions and performance. In this respect, the ability to carry out nonstationary, plurifrequential, and multivariate analyses of synchrony might lead to numerous significant advancements in our understanding of human social interaction.

**Declaration of Competing Interest**

None.

**Acknowledgement**

The authors would like to thank Birgitta Burger and Emily Carlson for their comments on prior versions of this manuscript and for supplying data associated with the dance movement task. They would also like to thank Marc Thompson for supplying mirror game data. This work was supported by funding from the Academy of Finland (project numbers 332331 and 314651).# 5  References